# Evidence for Lunar True Polar Wander, and a Past Low-Eccentricity, Synchronous Lunar Orbit

James Tuttle Keane[1] & Isamu Matsuyama[1]

**Corresponding Author:** J. T. Keane, Department of Planetary Science, Lunar and Planetary Laboratory, University of Arizona, Tucson, AZ 85721, United States. (jkeane@lpl.arizona.edu)

**Affiliations:** [1]Department of Planetary Science, Lunar and Planetary Laboratory, University of Arizona, Tucson, AZ 85721, United States.

**Key Points:**
- Lunar impact basins have a significant contribution to lunar inertia tensor.
- The Moon formed on a smaller, synchronous, low-eccentricity orbit.
- The formation of the South Pole-Aitken impact basin reoriented the Moon ~15°.

**Abstract:**
As first noted 200 years ago by Laplace, the Moon's rotational and tidal bulges are significantly larger than expected, given the Moon's present orbital and rotational state. This excess deformation has been ascribed to a fossil figure, frozen in when the Moon was closer to the Earth. However, the observed figure is only consistent with an eccentric and non-synchronous orbit, contrary to our understanding of the Moon's formation and evolution. Here, we show that lunar mascons and impact basins have a significant contribution to the observed lunar figure. Removing their contribution reveals a misaligned fossil figure consistent with an early epoch of true polar wander (driven by the formation of the South Pole-Aitken Basin) and an early low-eccentricity, synchronous lunar orbit. This new self-consistent model solves a long-standing problem in planetary science, and will inform future studies of the Moon's dynamical evolution and early dynamo.

# 1. Introduction

The long-wavelength lunar figure is triaxial, due to the combination of rotational deformation (which creates a rotational bulge) and tidal deformation (which creates a tidal bulge along the Earth-Moon axis). However, it has been known since Laplace [1878] that the observed deformation is significantly larger than the predicted deformation assuming hydrostatic equilibrium and the Moon's current orbital and rotational state (figure S1). This difference has been ascribed to the presence of a fossil figure, frozen in when the Moon was closer to the Earth and the rotational and tidal potentials were larger [Sedgwick, 1898; Jeffreys, 1915; Lambeck and Pullan, 1980]. More recent work by Garrick-Bethell et al. [2006] has shown that the observed deformation not only requires freezing the fossil figure while on an orbit with a smaller semimajor axis, but also with significant eccentricity. Matsuyama [2013] extended this work by taking into account the effect of finite lithospheric rigidity; however, the required initial orbital eccentricity remains large. For reasonable estimates of the rigidity of the lunar lithosphere, the fossil figure is consistent with initial semimajor axes and eccentricities ($a$, $e$) of (17.3 $R_\oplus$, 0.49) assuming synchronous rotation, and (18.2 $R_\oplus$, 0.16) or (20.1 $R_\oplus$, 0.60) assuming the lithosphere formed during a 3:2 spin-orbit resonance (Mercury's present spin-orbit state). Even the low-eccentricity 3:2 spin-orbit solutions favored by Garrick-Bethell et al. seem inconsistent with the favored model for the formation of the Moon from a circumplanetary debris disk, generated from the collision of a large protoplanet with the early Earth [Hartmann and Davis, 1975; Canup and Asphaug, 2001]. Collisional processing within a debris disk naturally yields satellites with near-zero eccentricity [Kokubo et al., 2000]. Tidal dissipation will also damp the Moon's eccentricity as it migrates outward. While transient periods of high eccentricity are possible if the Moon becomes trapped in the evection resonance [Touma and Wisdom, 1998], these high eccentricities would correspond to periods of extreme tidal heating, and it seems unlikely that the fossil figure would be able to freeze during such an event [Meyer et al., 2010].

Previous work investigating the lunar figure has assumed that the observed figure consists primarily of a fossil figure, or the relaxed remnant of such a figure. However, the formation and subsequent evolution of impact basins can have significant contributions to the lunar figure. Many large impact basins are associated with mascons – large uncompensated density anomalies that arise from the excavation, collapse, and isostatic adjustment of the impact crater, and (in some cases) subsequent volcanic flooding by denser mare basalt [Muller and Sjogren, 1968; Melosh et al. 2013]. Early investigations into the gravity contributions of nearside mascons noted a curious parallelism between the principal axes of the mascons and the principal axes of the Moon – suggesting that the mascons may be important for the lunar figure [O'Leary et al, 1968; Melosh, 1975a]. However, this hypothesis was subsequently dismissed, as the total mass of the nearside lunar mascons was insufficient to completely explain either the degree-2 figure or the magnitude and sign of the Moon's anomalously large degree-3 gravity field [Melosh, 1975a]. Non-mascon basins – such as the largest and oldest lunar impact basin, South Pole-Aitken (SPA) – may also contribute to the lunar figure and drive episodes of true polar wander [Melosh, 1975b; Ong and Melosh, 2010]. We revisit the contribution of mascons and impact basins (henceforth collectively referred to as mass anomalies) with the use of new high-precision, global gravity data from the GRAIL mission [Zuber et al., 2013].

The lunar figure can be uniquely quantified using the observed degree-2 spherical harmonic gravity coefficients: $C_{20}$ (-$J_2$), $C_{21}$, $C_{22}$, $S_{21}$, $S_{22}$. These degree-2 coefficients are directly related to the lunar inertia tensor [Lambeck, 1980]. In a coordinate system aligned with the principal moments of inertia, the figure is dependent only on the $C_{20}$ and $C_{22}$ terms, which quantify rotational and tidal deformation, respectively. As the lunar figure is determined solely from degree-2 coefficients, our goal is to determine the degree-2 coefficients associated with the lunar mass anomalies alone, and subtract their contribution from the observed figure (Sections 2.1 and 3). By isolating and removing the contribution of mass anomalies to the lunar figure, we are able to identify the Moon's true fossil figure, which holds important insight into the dynamical evolution of the Moon (Sections 2.2 and 3). It is important to note, when we refer to the "lunar figure," we are not directly referring to the Moon's physical, topographic shape. Rather, we are referring to the Moon's *gravitational* figure, as quantified by its degree-2 gravitational potential, or the equivalent inertia tensor and associated principal axes. The inertia tensor is what fundamentally governs the Moon's rotational dynamics. While topography can provide insight into the evolution of the Moon [e.g. Garrick-Bethell et al., 2014], topography alone cannot be related to the inertia tensor without some assumed model for the Moon's density structure.

## 2. Methods

### 2.1. The Contribution of Mass Anomalies to the Lunar Figure

Since impact basins and mascons are generally axisymmetric at long wavelength, we modeled the gravity field of each individual mass anomaly using a linear combination of concentric, uniform density spherical caps. A linear combination of caps with different radii and surface densities allows us to match the complicated radial "bulls-eye" gravity structure associated with many mascons [e.g. Melosh et al., 2013]. To determine the gravitational field associated with any individual cap, we first consider a cap centered on the lunar north pole. The gravity field of a circumpolar spherical cap is described entirely by zonal ($m=0$) spherical harmonic coefficients, $C_{l0}^{cap}$. Assuming a uniform surface density, $\sigma^{cap}$, and angular size, $\gamma^{cap}$, the zonal coefficients can be determined analytically:

$$C_{l0}^{cap} = \sigma^{cap} \frac{R_{\leftmoon}^2}{M_{\leftmoon}} \int_0^{\gamma^{cap}} \int_0^{2\pi} d\phi \, d\theta \sin\theta \; P_{l0}(\cos\theta) \qquad (1)$$

where $R_{\leftmoon}$ and $M_{\leftmoon}$ are the radius and mass of the Moon, $\delta$ is the Kronecker delta, $l$ and $m$ are the spherical harmonic degree and order, $P_{lm}$ is the unnormalized associated Legendre function [Wieczorek, 2007], and $\theta$ and $\phi$ are the colatitude and longitude. Both $\sigma^{cap}$ and $\gamma^{cap}$ are independent of the spherical harmonic degree and order. Using the spherical harmonic addition theorem [Arfken and Weber, 1995], we rotate the cap to be centered on any mass anomaly of interest:

$$\begin{bmatrix} C_{lm}^{cap} \\ S_{lm}^{Cap} \end{bmatrix} = C_{l0}^{cap} \, (2 - \delta_{m0}) \frac{(l-m)!}{(l+m)!} P_{lm}(\cos\theta^{cap}) \begin{bmatrix} \cos(m\phi^{cap}) \\ \sin(m\phi^{cap}) \end{bmatrix} \qquad (2)$$

where $\theta^{cap}$ and $\phi^{cap}$ are the colatitude and longitude of the mass anomaly. From these spherical harmonic coefficients we then evaluate the gravitational potential, $U^{cap}$, and free-air anomaly, $F^{cap}$,

$$U^{cap}(\theta, \phi) = \frac{GM_{\mathbb{C}}}{r} \sum_{l=0}^{l_{max}} \sum_{m=0}^{l} \left(\frac{R_{\mathbb{C}}}{r}\right)^2 [\, C_{lm}^{cap} P_{lm}(\cos\theta)\cos(m\phi) \\ + S_{lm}^{cap} P_{lm}(\cos\theta)\sin(m\phi)] \quad (3)$$

$$F^{cap}(\theta, \phi) = \frac{GM_{\mathbb{C}}}{r^2} \sum_{l=0}^{l_{max}} \sum_{m=0}^{l} \left(\frac{R_{\mathbb{C}}}{r}\right)^2 (l+1)[\, C_{lm}^{cap} P_{lm}(\cos\theta)\cos(m\phi) + S_{lm}^{cap} P_{lm}(\cos\theta)\sin(m\phi)] \quad (4)$$

associated with the cap as a function of selenographic colatitude, $\theta$, and longitude, $\phi$, for a fixed radius ($r=R_{\mathbb{C}}$). Since the cap's spherical harmonics, gravitational potential and free-air anomaly scale linearly with the cap's surface density, $\sigma^{cap}$, it is convenient to describe these quantities per unit surface density: $F^{cap'}$, $U^{cap'}$, $C_{lm}^{cap'}$, $S_{lm}^{cap'}$ (where $F^{cap'}=F^{cap}/\sigma^{cap}$, etc.). These quantities are henceforth referred to as *specific* free-air anomaly, *specific* gravitational potential, etc.

For each mass anomaly, we generated a set of spherical caps spanning a range of radii encompassing the entire gravitational structure associated with each anomaly. For mascons, this included caps from within the central free-air anomaly high to outside the entire "bull's-eye" structure [e.g. Melosh et al., 2013]. For SPA, this included caps extending from within the central topographic low to beyond the spherically symmetric near-equatorial topographic high thought to be associated with its global ejecta blanket [Ong and Melosh, 2010]. The number of caps (and cap spacing) used for any individual mass anomaly is set such that the cap size/spacing is commensurate with the spatial resolution of our maximum spherical harmonic degree/order ($\theta \approx 360°/(2l_{max})$). For small impact basins (e.g. Grimaldi and D'Alembert), this corresponds to $\geq 2$ caps; large impact basins (e.g. Imbrium and Orientale), $\geq 6$ caps; SPA $\geq 20$ caps. Thus, for a typical inverse, the total number of spherical caps is >100. For each cap we evaluated the specific gravitational potential, $U^{cap'}$, and specific free-air anomaly, $F^{cap'}$, over a global Cartesian colatitude/longitude grid, from spherical harmonic degree 3–40. To determine the best-fitting surface densities for each spherical cap ($\sigma^{cap}$), we fit $U^{cap'}$ and $F^{cap'}$ to the GRAIL derived gravitational potential, $U^{Obs}$, and specific free-air anomaly, $F^{Obs}$, evaluated over the same global Cartesian grid from spherical harmonic degree 3–40. This fitting is performed from degree-3 and up in order to prevent directly fitting any underlying fossil figure (which is solely degree-2). The maximum degree is arbitrary, and does not significantly affect our results. This linear relationship can be written as:

$$\begin{bmatrix} F_1^{Obs} \\ F_2^{Obs} \\ \vdots \\ F_N^{Obs} \\ U_1^{Obs} \\ U_2^{Obs} \\ \vdots \\ U_N^{Obs} \end{bmatrix} = \sigma_1 \begin{bmatrix} F_1^{cap',1} \\ F_2^{cap',1} \\ \vdots \\ F_N^{cap',1} \\ U_1^{cap',1} \\ U_2^{cap',1} \\ \vdots \\ U_N^{cap',1} \end{bmatrix} + \sigma_2 \begin{bmatrix} F_1^{cap',2} \\ F_2^{cap',2} \\ \vdots \\ F_N^{cap',2} \\ U_1^{cap',2} \\ U_2^{cap',2} \\ \vdots \\ U_N^{cap',2} \end{bmatrix} + \cdots + \sigma_M \begin{bmatrix} F_1^{cap',M} \\ F_2^{cap',M} \\ \vdots \\ F_N^{cap',M} \\ U_1^{cap',M} \\ U_2^{cap',M} \\ \vdots \\ U_N^{cap',M} \end{bmatrix} \quad (5)$$

where the subscripts (1→N) specify the latitude/longitude index, and the sub/superscripts

(1→M) specify the cap. This relationship can be simplified in matrix notation:
$$\boldsymbol{d} = \boldsymbol{G}\boldsymbol{\sigma} \tag{6}$$
where $\boldsymbol{d}$ is an 2N×1 column vector containing the observed free-air anomaly and gravitational potential; $\boldsymbol{\sigma}$ is an M×1 column vector of the surface densities for each of the M caps; and $\boldsymbol{G}$ is the 2N×M data kernel, containing the specific free-air anomalies and specific gravitational potentials for each cap. The best-fit linear combination of surface densities for the array of caps can be calculated using weighted least squares [e.g. Menke, 2012]:
$$\boldsymbol{\sigma}_{WLS} = \{[\boldsymbol{G}^T \boldsymbol{W} \boldsymbol{G}]^{-1} \boldsymbol{G}^T \boldsymbol{W}\} \boldsymbol{d} \tag{7}$$
where the term in the curly brackets is the inverse operator, $\boldsymbol{G}^{INV}$, and $\boldsymbol{W}$ is an 2N×2N, diagonal weighting matrix. The diagonal elements of $\boldsymbol{W}$ correspond to the surface area of each colatitude/longitude point. This weighting acts to prevent biases toward polar regions where colatitude/longitude points converge in our Cartesian grid. To ensure accuracy and stability, we evaluate the model resolution matrix, $\boldsymbol{R}=\boldsymbol{G}^{INV}\boldsymbol{G}$, and condition number, and only considered solutions where each model parameter is uniquely resolved (in which case, $\boldsymbol{R}=\boldsymbol{I}$, where $\boldsymbol{I}$ is the identity matrix; $\boldsymbol{G}^{INV}$ is non-singular) and the result is computationally stable. While we performed simultaneous fits using both gravitational potential and free-air anomaly, these two quantities are intricately related (equations 3 and 4). Gravitational potential highlights long-wavelength features (e.g. the fossil figure and SPA) while free-air anomaly highlights short-wavelength features (e.g. mascons). Using one or the other does not significantly change the solution.

Once the best-fit surface densities for the set of spherical caps are evaluated from degree 3–40, it is possible to *directly* derive the degree-2 gravity coefficients associated with any mass anomaly. As shown in equations (1) and (2), the spherical harmonic gravity coefficients for any cap ($C_{lm}^{cap}$ and $S_{lm}^{cap}$) scale linearly with the assumed surface density of that cap (i.e. $C_{lm}^{cap}=\sigma^{cap}C_{lm}^{cap'}$). Thus, the spherical harmonic gravity coefficients for the entire mass anomaly are simply the linear combination of specific gravity coefficients times their best-fit surface densities:

$$\begin{bmatrix} C_{00} \\ C_{10} \\ C_{11} \\ C_{20} \\ C_{21} \\ \vdots \end{bmatrix} = \sigma_{WLS,1} \begin{bmatrix} C_{00}^{cap',1} \\ C_{10}^{cap',1} \\ C_{11}^{cap',1} \\ C_{20}^{cap',1} \\ C_{21}^{cap',1} \\ \vdots \end{bmatrix} + \sigma_{WLS,2} \begin{bmatrix} C_{00}^{cap',2} \\ C_{10}^{cap',2} \\ C_{11}^{cap',2} \\ C_{20}^{cap',2} \\ C_{21}^{cap',2} \\ \vdots \end{bmatrix} + \cdots + \sigma_{WLS,M} \begin{bmatrix} C_{00}^{cap',M} \\ C_{10}^{cap',M} \\ C_{11}^{cap',M} \\ C_{20}^{cap',M} \\ C_{21}^{cap',M} \\ \vdots \end{bmatrix} \tag{8}$$

where $\sigma_{WLS,1} \rightarrow \sigma_{WLS,M}$ are the best fit surface densities for spherical caps 1→M (the components of the vector $\boldsymbol{\sigma}_{WLS}$ in equation (7)). As evident in equation (8), spherical caps also have some power in degree-0 and degree-1. Degree-0 corresponds to the addition or subtraction of mass from the Moon by the cap. Degree-1 corresponds to an offset of the center-of-mass. In a center-of-mass reference frame, degree-1 spherical harmonic gravity coefficients are zero. Since these coefficients are generally evaluated in center-of-mass reference frame, it is important to identify and correct for any offset resulting from our removal of the mass anomalies. In general, our solutions yield center-of-mass offsets of ~1 km, which result in negligible errors (<1%) in the degree-2 gravity coefficients. Thus, we ignore the effects of the center-of-mass offset. A simple demonstration of this mass anomaly fitting for a SPA-like mass anomaly is included in figure (S2). It is worth noting that this 1 km center-of-mass offset is in the direction of SPA and the lunar farside. This reduces the well-known center-of-mass/center-of-figure offset [~2 km, Smith et al. 1997], and will be investigated in future work.

The uncertainties in the best-fit surface densities for any set of spherical caps can be propagated in the usual way [e.g. Tellinghuisen, 2001]:

$$unc_{\sigma_{WLS}} = \sigma_{WLS}{}^T \left\{ G^{INV} G^{INV^T} \right\} \sigma_{WLS} \qquad (9)$$

where the term in the brackets is the covariance matrix. In general, the uncertainties in the best-fit surface densities from any individual fit are negligible. A more significant source of uncertainty in the best-fit surface densities (and the resulting spherical harmonics) is the number and spacing of the caps and the choice of latitude/longitude grid. To quantify these uncertainties, we experimented with varying these quantities and repeating the inverse. Varying these parameters within reasonable tolerances (grid and cap spacing between 1.5° and 6°; number of spherical caps varying by a factor of 4) yielded uncertainties in the fossil figure of $\pm 3 \times 10^{-6}$ (2%) and $\pm 4 \times 10^{-6}$ (10%) for $C_{20}$ and $C_{22}$, respectively. An additional source of uncertainty is associated with the alignment of the caps with the mass anomalies. We quantified the effects of misalignments through investigation of the forward problem. We superimposed a SPA-like mass anomaly onto a synthetic fossil figure, and attempted to retrieve the input fossil figure with our spherical cap inverse method (figure S2). By randomly shifting the center of the caps with respect to the true center of the mass anomaly, we quantified the accuracy of the inverse method as a function of the magnitude of the misalignment. Assuming we are able to accurately place the centers of mass anomalies within ~5°, results in an added uncertainty of $\pm 11 \times 10^{-6}$ (8%) and $\pm 2 \times 10^{6}$ (2%) for $C_{20}$ and $C_{22}$, for SPA-like mass anomalies (figure S3). The total uncertainty is found from adding these individual uncertainties in quadrature.

It is important to contrast our methodology with that of Garrick-Bethell et al. [2014]. In their work, Garrick-Bethell et al. attempt to fit the underlying fossil figure by fitting the degree-2 gravitational potential of the Moon *outside* of lunar impact basins. However, this methodology significantly underestimates the contribution of mass anomalies. The degree-2 gravity field associated with any mass anomaly is, by definition, global. To illustrate this, consider the null solution where the Moon possesses *no* fossil figure ($C_{20} = C_{22} = 0$), but has single, large, mass anomaly such as South Pole-Aitken (figure S4). While masking the basin from the global gravitational potential will hide the high-order gravity power associated with anomaly (figure S4a), it will not hide the majority of the degree-2 gravity field associated with the basin (figure S4b). Thus, a fit to the degree-2 field outside of the basin will still retain a significant fraction of the degree-2 field associated with the basin. This is why Garrick-Bethell et al. find that impact basins have negligible contribution to degree-2 gravity. A comparison of these two methodologies is shown in figure S5.

## 2.2. Orbital Solutions for the Fossil Figure

Once we subtract the contribution of mass anomalies from the lunar figure, we arrive at the true lunar fossil figure. This figure represents the relaxed remnant of a fossil rotational and tidal bulge, frozen in when the lunar lithosphere became rigid enough to support long-term deformations, as well as some small present-day hydrostatic component. The exact nature of this figure can be calculated directly using secular Love number theory.

Following Matsuyama and Nimmo [2009], and Matsuyama [2013], we quantify the lunar figure in terms of principal moments of inertia, $A<B<C$, and their deviations from the mean moment of inertia, $I$. In this notation, the observed lunar figure $[A^{OBS}, B^{OBS}, C^{OBS}]$ is an amalgamation of a primordial figure from when the elastic lithosphere forms $[A^*, B^*, C^*]$, the subsequent deformation of the moon as it migrates outward through different rotational and tidal potentials $[A^{DEF}, B^{DEF}, C^{DEF}]$, and any contribution from mass anomalies $[A^{MA}, B^{MA}, C^{MA}]$:

$$[A^{OBS}, B^{OBS}, C^{OBS}] = [A^*, B^*, C^*] + [A^{DEF}, B^{DEF}, C^{DEF}] + [A^{MA}, B^{MA}, C^{MA}] \quad (10)$$

Any set of principal moments of inertia can be uniquely determined by the corresponding set of degree-2 spherical harmonic gravity coefficients, by way of the inertia tensor, $I$ [Lambeck, 1980]:

$$I_{ij} = I\delta_{ij} + M_{\mathbb{C}} R_{\mathbb{C}}^2 \begin{bmatrix} \frac{1}{3}C_{20} - 2C_{22} & -2S_{22} & -C_{21} \\ -2S_{22} & \frac{1}{3}C_{20} + 2C_{22} & -S_{21} \\ -C_{21} & -S_{21} & \frac{2}{3}C_{20} \end{bmatrix} \quad (11)$$

where $\delta_{ij}$ is the Kronecker delta, and $I$ is the spherically symmetric, mean moment of inertia ($I = 8\pi/3 \int_0^{R_{\mathbb{C}}} dr\, r^4 \rho$). The principal moments are found by diagonalizing the inertia tensor.

The moments of inertia of the Moon at the time its elastic lithosphere formed, $[A^*, B^*, C^*]$, can be written [Matsuyama 2013]:

$$\begin{aligned}
\frac{C^*}{M_{\mathbb{C}} R_{\mathbb{C}}^2} &= \frac{I}{M_{\mathbb{C}} R_{\mathbb{C}}^2} + k_2^* \frac{1}{9} \frac{M_\oplus R_{\mathbb{C}}^3}{M_{\mathbb{C}} a_*^3} \left[ 2p_*^2 + 3(1-e_*^2)^{-\frac{3}{2}} \right] \\
\frac{B^*}{M_{\mathbb{C}} R_{\mathbb{C}}^2} &= \frac{I}{M_{\mathbb{C}} R_{\mathbb{C}}^2} + k_2^* \frac{1}{18} \frac{M_\oplus R_{\mathbb{C}}^3}{M_{\mathbb{C}} a_*^3} \left[ -2p_*^2 - 3(1-e_*^2)^{-\frac{3}{2}} + 9H(p_*, e_*) \right] \\
\frac{A^*}{M_{\mathbb{C}} R_{\mathbb{C}}^2} &= \frac{I}{M_{\mathbb{C}} R_{\mathbb{C}}^2} + k_2^* \frac{1}{18} \frac{M_\oplus R_{\mathbb{C}}^3}{M_{\mathbb{C}} a_*^3} \left[ -2p_*^2 - 3(1-e_*^2)^{-\frac{3}{2}} - 9H(p_*, e_*) \right]
\end{aligned} \quad (12)$$

where $M_\oplus$ is the mass of the Earth, and $a_*$, $e_*$ and $p_*$ are the semimajor axis, eccentricity, and spin-orbit ratio at the time the fossil figure is established. $k_2^*$ is the secular degree-2 tidal Love number for the case *without* an elastic lithosphere. We evaluated $k_2^*$ using a nominal 3-layer interior structure model consisting of a core, mantle and elastic lithosphere of varying thickness [see: Sabadini and Vermeersen, 2004; Matsuyama, 2013]. We assume that the Moon lacked an elastic lithosphere at the time the fossil figure froze, resulting in $k_2^*=1.44$. This value is very nearly the same as the case for a homogeneous strengthless Moon, $k_2^*=1.5$ (as expected because the Moon is nearly homogeneous). Our results are not strongly dependent on the value of $k_2^*$. $H(p_*,e_*)$ are Hansen coefficients, which describe the time-averaged potential for a satellite as a function of spin-orbit ratio and eccentricity [Goldreich, 1966]. To eighth-order in eccentricity:

$$\begin{aligned}
H(1/1, e) &= 1 - \frac{5}{2}e^2 + \frac{13}{16}e^4 + \frac{35}{288}e^6 + \frac{5}{576}e^8 + \cdots \\
H(3/2, e) &= \frac{7}{2}e - \frac{123}{16}e^3 + \frac{489}{128}e^5 + \frac{1763}{2048}e^7 + \cdots \\
H(2/1, e) &= \frac{17}{2}e^2 - \frac{115}{6}e^4 + \frac{601}{48}e^6 + \frac{1423}{360}e^8 + \cdots
\end{aligned} \quad (13)$$

As the Moon migrates away from the Earth, it deforms in response to changes in the rotational and tidal potentials [Matsuyama and Nimmo, 2006; Matsuyama, 2013]:

$$\frac{C^{DEF}}{M_{\mathbb{C}} R_{\mathbb{C}}^2} = k_2 \frac{1}{9} \frac{M_\oplus}{M_{\mathbb{C}}} \left\{ \frac{R_{\mathbb{C}}^3}{a^3} \left[ 2p + 3(1-e^2)^{-\frac{3}{2}} \right] - \frac{r^3}{a_*^3} \left[ 2p_*^2 + 3(1-e_*^2)^{-\frac{3}{2}} \right] \right\}$$

$$\frac{B^{DEF}}{M_{\mathbb{C}} R_{\mathbb{C}}^2} = k_2 \frac{1}{18} \frac{M_\oplus}{M_{\mathbb{C}}} \left\{ \frac{R_{\mathbb{C}}^3}{a^3} \left[ -2p^2 - 3(1-e^2)^{-\frac{3}{2}} + 9H(p,e) \right] \right.$$
$$\left. - \frac{R_{\mathbb{C}}^3}{a_*^3} \left[ -2p_*^2 - 3(1-e_*^2)^{-\frac{3}{2}} + 9H(p_*,e_*) \right] \right\} \quad (14)$$

$$\frac{A^{DEF}}{M_{\mathbb{C}} R_{\mathbb{C}}^2} = k_2 \frac{1}{18} \frac{M_\oplus}{M_{\mathbb{C}}} \left\{ \frac{R_{\mathbb{C}}^3}{a^3} \left[ -2p^2 - 3(1-e^2)^{-\frac{3}{2}} - 9H(p,e) \right] \right.$$
$$\left. - \frac{R_{\mathbb{C}}^3}{a_*^3} \left[ -2p_*^2 - 3(1-e_*^2)^{-\frac{3}{2}} - 9H(p_*,e_*) \right] \right\}$$

where $a$, $e$, and $p$, are the present semimajor axis, eccentricity, and spin-orbit ratio of the Moon. $k_2$ is the secular degree-2 tidal love number for the case *with* an elastic lithosphere. In the limit of a strengthless lithosphere ($k_2=k_2^*$) and the lithosphere would be incapable of supporting any fossil figure. For a lithosphere with infinite rigidity ($k_2=0$), this deformation term vanishes, and the fossil figure would be completely preserved (as is the case for Garrick-Bethell et al. [2006]). For this work, we follow Matsuyama [2013] and consider 50 km ($k_2=0.86$) and 25 km ($k_2=1.08$) thick elastic lithospheres.

## 3. Results

Using our spherical cap inverse method (section 2.1), we modeled the gravity fields of the 31 largest lunar mass anomalies. Figures 1c and 1d illustrate one example fit to the gravitational potential for these mass anomalies. The single most important mass anomaly is the SPA basin and its associated ejecta blanket. Analysis of the resulting spherical harmonics (table S1) reveals that SPA accounts for 96% of the total degree-2 power from the mass anomalies. Additionally, SPA accounts for 82% of the total degree-3 power associated with the mass anomalies, and 80% of the *observed* degree-3 power. This confirms the early predictions of Melosh [1975a], who first used degree-3 gravity to hypothesize that some far-side mass anomaly was controlling the Moon's present orientation. In contrast, the nearside flood basalt filled impact basins have negligible contributions to degree-2 (<1%) and degree-3 (5%).

Figure 1e illustrates the gravitational field that results when we subtract the contribution of lunar mass anomalies (figure 1c) from the observed lunar gravity field (figure 1a). The most notable aspect of this mass-anomaly corrected fossil figure is the obvious misalignment between its principal axes and the principal axes of the present-day Moon. For a tidally deformed body like the Moon, the minimum energy state corresponds to a configuration where the rotation and tidal axes are aligned with the maximum and minimum principal axes of the inertia tensor, respectively. This is the case for the present-day Moon. The observed misalignment between the principal axes of the corrected fossil figure and the present figure is the signature of reorientation of the rotational and tidal axes relative to the surface, or true polar wander. By converting the mass-anomaly corrected degree-2 gravity coefficients to the corresponding inertia tensor (eq. 11) we find the fossil figure has reoriented ~15°. The sum of this misaligned fossil figure (figure 1e) and the mass anomalies (figure 1c) produces the present inertia tensor with its principal axes aligned with the rotational and tidal axes (figure 1a). The primordial tidal axis (the minimum principal moment of inertia of the fossil figure) is located at (14.3±1.4°N, 1.9±0.9°E); the primordial rotation axis (the maximum principal moment of inertia of the fossil figure) is located at (71.7±2.0°N, 221.6±2.1°E). This reorientation is consistent with SPA being a negative-mass anomaly at degree-2, originally located near the equator and Earth-Moon tidal axis. After the formation of SPA, the Moon reoriented to place this negative anomaly closer to the lunar south pole. Since the fossil figure necessarily formed with its principal axes aligned with the rotational and tidal axes, we diagonalized the inertia tensor associated with the mass–anomaly corrected fossil figure and recalculated the degree-2 gravity coefficients in a principal axes coordinate system (table S1).

Having corrected for mass anomalies and true polar wander, we now consider possible orbital solutions that can reproduce the corrected fossil figure using secular Love number theory (section 2.2). We evaluated the predicted fossil figure (as quantified by principal moments of inertia, or degree-2 gravity coefficients: $C_{20}$ and $C_{22}$) for a range of initial semimajor axes, eccentricities, spin-orbit resonances, and considering 25 km and 50 km thick elastic lithospheres, consistent with previous estimates from lunar gravity and topography data [although the exact nature of the elastic lithosphere at this time is still debated; Arkani-Hamed, 1998; Sugano and Heki, 2004; Crosby and McKenzie, 2005; Audet and Johnson, 2005]. We then compared these fossil figure predictions to our mass anomaly and true polar wander corrected fossil figure. Figure 2 illustrates the semimajor axis and eccentricity solutions as a function of the predicted fossil figure's degree-2 gravity coefficients, $C_{20}$ and $C_{22}$, for a 50 km thick elastic lithosphere for 1:1 (figure 2a) and 3:2 spin-orbit resonances (figures 2b and 2c). In these plots, regions overlain by contours indicate orbital solutions that yield self-consistent fossil figures; regions devoid of contours indicate that that particular fossil figure cannot be self-consistently created from that particular combination of spin-orbit resonance and elastic lithosphere thickness. Figure S7 illustrates the same results in the style of Garrick-Bethell et al. [2006] and Matsuyama [2013].

If we assume that the elastic lithosphere formed during synchronous rotation, our corrected fossil figure is consistent with an initial semimajor axis and eccentricity ($a$, $e$) of (17.8±0.2 $R_\oplus$, 0.21±0.05) for a 50 km thick elastic lithosphere (figure 2a), and (15.2±0.2 $R_\oplus$, 0.22±0.05) for a 25 km thick elastic lithosphere (figure S8a). These low-eccentricity solutions stand in stark contrast to the orbital solutions of Garrick-Bethell et al. [2006] and Matsuyama [2013], who both required higher-eccentricity orbits at the time that the fossil figure froze in ($e$~0.5). Our lower-eccentricity solutions are more consistent with the canonical model for the formation and evolution of the Moon. The $C_{20}/C_{22}$-ratio of our corrected fossil figure is very close to the hydrostatic, zero-eccentricity synchronous orbit ratio of -10/3 predicted by Laplace [1878]: $C_{20}/C_{22}$=-4.03±0.28. The hydrostatic solution for a satellite with e=0.2 is $C_{20}/C_{22}$~-4 [Matsuyama and Nimmo, 2009]. While our corrected fossil figure does not require significant primordial eccentricity, this does not preclude earlier epochs of high eccentricity that may naturally result from the evection resonance; we simply do not require the Moon to freeze its figure during these epochs.

In contrast with Garrick-Bethell et al. [2006] and Matsuyama [2013], we find that the corrected fossil figure is inconsistent with initial 3:2 or 2:1 spin-orbit resonances (figures 2b, 2c,

S8b, S8c, and S9). While the Moon may have passed through these higher-order spin-orbit resonances [Goldreich and Peale, 1966], our work suggests that the Moon was synchronously locked by the time the fossil figure froze in.

## 4. Summary

We have developed a novel technique for isolating the degree-2 gravitational field of lunar mass anomalies. By subtracting the contribution of mass anomalies, we determine the Moon's true fossil figure. This mass anomaly corrected fossil figure is misaligned with the present day principal axes of the Moon, suggesting past true polar wander driven primarily by the formation of SPA. Using secular Love number theory, we demonstrate that this corrected lunar figure formed when the Moon was in a synchronous orbit with a semimajor axes of 15-17 $R_\oplus$ and an eccentricity of ~0.2. While not negligible, this eccentricity is considerably lower than previous estimates [Garrick-Bethell et al., 2006; Matsuyama 2013]. We exclude the possibility that the fossil figure formed during higher-order 3:2 or 2:1 spin-orbit resonances.

Figure 3 illustrates our new picture for the formation and evolution of the lunar figure. After the formation of the Moon (figure 3a), the Moon migrated outward under the action of tides (figure 3b). After any epoch of non-synchronous rotation or high-eccentricity (e.g. the evection resonance), the Moon settled into a low-eccentricity, synchronous orbit, and cooled sufficiently, forming an elastic lithosphere capable of supporting the fossil figure (figure 3c). Later, the SPA basin-forming impact occurred, and reoriented the Moon by ~15° (figure 3d and 3e). The amount of reorientation is smaller than those inferred by magnetization estimates of pre-Nectarian paleopoles, although the direction of reorientation is similar [Runcorn, 1983; Takahashi et al., 2014]. The remaining large lunar impact basins and mare flood basalts formed after the formation of SPA (figure 3f). Curiously, the majority of the mare flood basalts (and the Procellellarum KREEP terrain) lie near the primordial tidal axis that was displaced northward by the formation of SPA. This correlation may be the result of tectonics driven from the reorientation and despinning of the Moon.

**Acknowledgments:**

J. T. K. and I. M. acknowledge support from the National Aeronautics and Space Administration Lunar Advanced Science and Exploration Research (LASER) program (NNX12AI98G) and the GRAIL Guest Science Program (NNX12AL09G). J. T. K. thanks R. M. Richardson for resourceful discussions regarding inverse methods. The GRAIL data used in this study are available from the Geosciences Node of the Planetary Data System: http://geo.pds.nasa.gov/missions/grail/default.htm.

**References:**

Arfken, G. & Weber, H. (1995), *Mathematical Methods for Physicists, 4$^{th}$ Edition,* Academic Press, San Francisco, CA.

Arkani-Hamed, J. (1998), The lunar mascons revisited. *J. Geophys. Res.,* **103**, 3709-3739, doi: 10.1029/97JE02815.

Audet, P. & Johnson, C. L. (2011), Lithospheric structure of the Moon and the correlation with deep moonquake source regions. *Lunar Planet. Sci. Conf.*, **42**, 1742.

Canup, R. M. & Asphaug, E. (2001), Origin of the Moon in a giant impact near the end of the Earth's formation. *Nature*, **412**, 708-712.

Crosby, A. & McKenzie, D. (2005), Measurements of the elastic thickness under ancient lunar terrain. *Icarus*, **173**, 100-107, doi: 10.1016/j.icarus.2004.07.017.

Garrick-Bethell, I. Wisdom, J. & Zuber, M. T. (2006), Evidence for a past high-eccentricity lunar orbit. *Science*, **313**, 652-655, doi: 10.1126/science.1128237.

Garrick-Bethell, I., Perera, V., Nimmo F., Zuber, M. T. (2014), The tidal-rotational shape of the Moon and evidence for polar wander. *Nature,* **512***,* 181-184, doi: 10.1038/nature13639.

Goldreich, P. (1966), History of the lunar orbit. *Rev. Geophys. Space Phys,*. **4**, 411-439, doi: 10.1029/RG004i004p00411.

Goldreich, P. & Peale, S. (1966), Spin-orbit coupling in the solar system. *Astronomical Journal*, **71**, 425-437, doi: 10.1086/109947.

Hartmann, W. K. & Davis, D. R. (1975) Satellite-sized planetesimals and lunar origin. *Icarus*, **24**, 504-514, doi: 10.1016/0019-1035(75)90070-6.

Jeffreys, H. (1915), Certain hypothesis as to the internal structure of the Earth and Moon. *Mem. R. Astron. Soc.*, **60**, 187-217.

Kokubo, E., Ida, S., & Makino, J. (2000), Evolution of a circumterrestrial disk and formation of a single moon. *Icarus*, **148**, 419-436, doi: 10.1006/icar.2000.6496.

Lambeck, K. (1980), *The Earth's Variable Rotation: Geophysical Causes and Consequences*, Cambridge Univ. Press, New York, NY.

Lambeck, K. & Pullan, S. (1980), The lunar fossil bulge hypothesis revisited. *Phys. Earth Planet. Interiors*, **22**, 29-35, doi: 10.1016/0031-9201(80)90097-7.

Laplace, P.-S. (1878), *Oeuvres comptetes de Laplace*, Gauthiers-Villars, Paris.

Matsuyama, I. (2013), Fossil figure contribution to the lunar figure. *Icarus*, **222**, 411-414, doi: 10.1016/j.icarus.2012.10.025.

Matsuyama, I. & Nimmo, F. (2009), Gravity and tectonic patterns of Mercury: effect of tidal deformation, spin-orbit resonance, nonzero eccentricity, despinning, and reorientation. *Journal Geophys. Res.*, **114**, E01010, doi: 10.1029/2008JE003252.

Melosh, H. J. (1975a), Mascons and the Moon's orientation. *Earth and Planet. Sci. Lett.*, **25**, 322-326, doi: 10.1016/0012-821X(75)90248-4.

Melosh, H. J. (1975b), Large impact craters and the Moon's orientation. *Earth and Planet. Sci. Lett.*, **26**, 353-360, doi: 10.1016/0012-821X(75)90011-4.

Melosh, H. J. et al. (2013), The origin of lunar mascon basins. *Science*, **340**, 1552-1555, doi: 10.1126/science.1235768.

Menke, W. (2012), *Geophysical Data Analysis: Discrete Inverse Theory, 3$^{rd}$ Edition,* Academic Press, San Diego, CA.

Meyer, J., Elkins-Tanton, L., Wisdom, J. (2010), Coupled thermal-orbital evolution of the early Moon. *Icarus*, **208**, 1-10, doi: 10.1016/j.icarus.2010.01.029.

Muller, P. M. & Sjogren, W. L. (1968), Mascons: lunar mass concentrations. *Science*, **161**, 680-684, doi: 10.1126/science.161.3842.680.

O'Leary, B. T., Campbell, M. J., & Sagan C. (1969), Lunar and Planetary Mass Concentrations. *Science*, **165**, 651-657, doi: 10.1038/2201309a0.

Ong, L. & Melosh, H. J. (2010), Reorientation of the Moon due to SPA basin ejecta. *Lunar Planet. Sci. Conf.*, **41**, 1363.

Runcorn, S. K. (1983), Lunar magnetism, polar displacements and primeval satellites in the earth-moon system. *Nature*, **304**, 589-596, doi: 10.1038/304589a0.

Sabadini, R. & Vermeersen, B. (2004), *Global Dynamics of the Earth: Applications of Normal Mode Relaxation Theory to Solid-Earth Geophysics*, Kluwer Academic Publishers, Dordrecht, the Netherlands.

Sedgwick, W. F. (1898), On the oscillations of a heterogeneous compressible liquid sphere and the genesis of the Moon; and the figure of the Moon. *Messenger Math.*, **27**, 159-173.

Smith, D. E, Zuber, M. T., Neumann, G. A. (1997), Topography of the Moon from the Clementine lidar. *JGR*, **102**, 1591-1611, doi: 10.1029/96JE02940.

Sugano, T. & Heki, K. (2004), Isostasy of the Moon from high-resolution gravity and topography data: implications for its thermal history. *Geophys. Res. Lett.*, **31**, L24703, doi: 10.1029/2004GL022059.

Takahashi, F., et al. (2014), Reorientation of the early lunar pole. *Nature Geoscience*, **7**, 409-412, doi: 10.1038/NGEO2150.

Tellinghuisen, J. (2001), Statistical Error Propogation, *J. Phys. Chem. A.*, 105, 3917-2921.

Touma, J. & Wisdom J. (1998), Resonances in the early evolution of the Earth-Moon system. *AJ*, **115**, 1653-1663, doi: 10.1086/300312.

Wieczorek, M. A. (2007), Gravity and topography of the terrestrial planets. *Treatise on Geophysics*, **10**, 165-206, doi: 10.1016/B978-044452748-6/00156-5.

Zuber, M. T. et al. (2013), Gravity field of the Moon from the Gravity Recovery and Interior Laboratory (GRAIL) mission. *Science*, **339**, 668-671, doi: 10.1126/science.1231507.

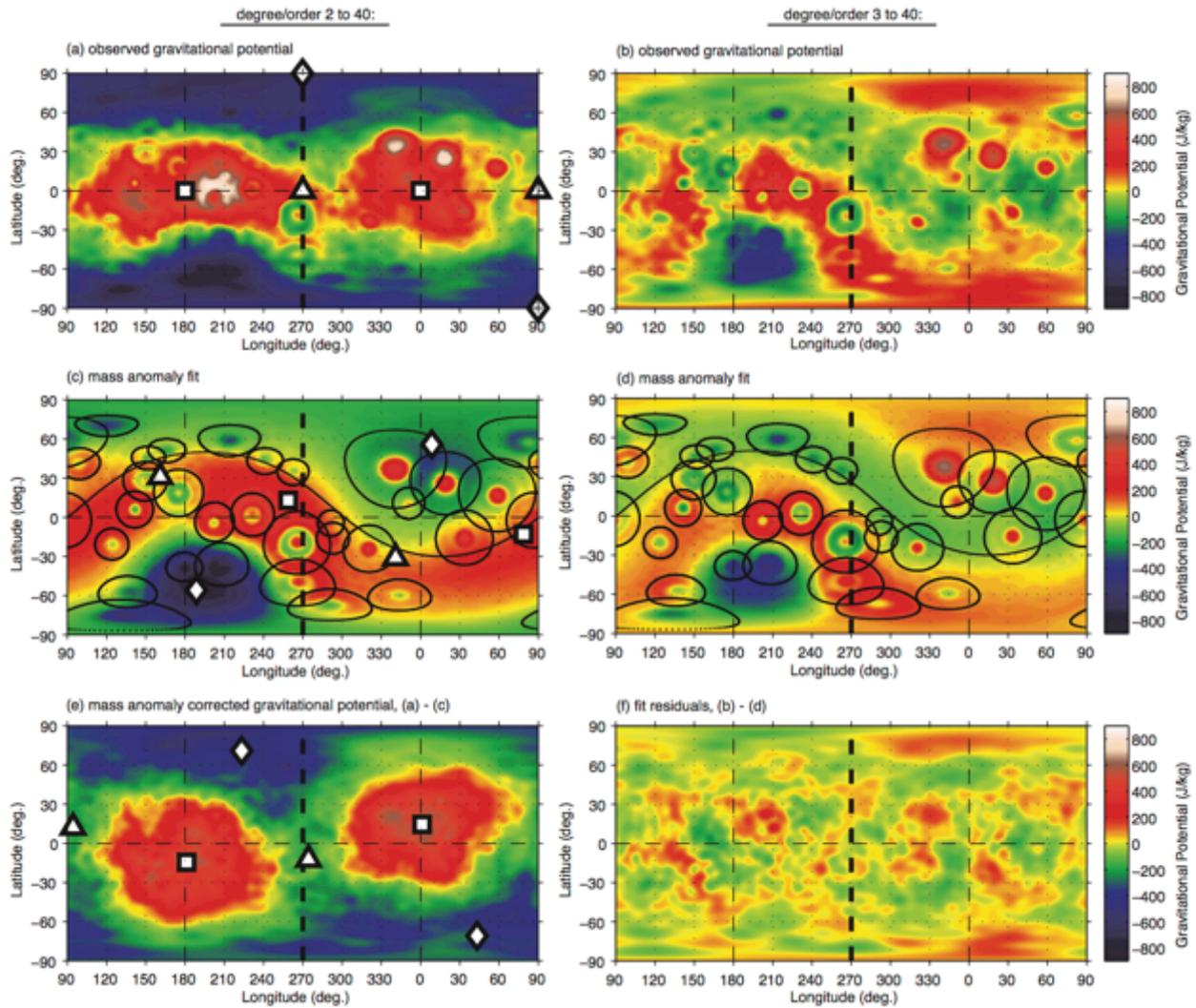

**Figure 1.** The observed, best-fit model, and the mass anomaly corrected gravitational potential of the Moon. (a) Observed gravitational potential of the Moon, from degree 2–40. Principal moments of inertia are indicated with squares (minimum principal moment, A, along tidal axis), triangles (intermediate principal moment, B), and diamonds (maximum principal moment, C, along rotational axis). Mass anomalies are outlined. (b) Observed gravitational potential of the Moon, from degree 3–40. The gravity contributions of mass anomalies are fit from degree-3 and up to prevent directly fitting any underlying fossil figure. (c and d) Example fit to gravity signature of the mass anomalies. (e and f) The residuals after subtracting out the mass anomaly model from the observed potential. The residuals from degree 2–40 (e) show the mass-anomaly corrected fossil figure, while residuals from degree 3-40 (f) are fit residuals. Figure S6 shows the same results in terms of free-air anomaly.

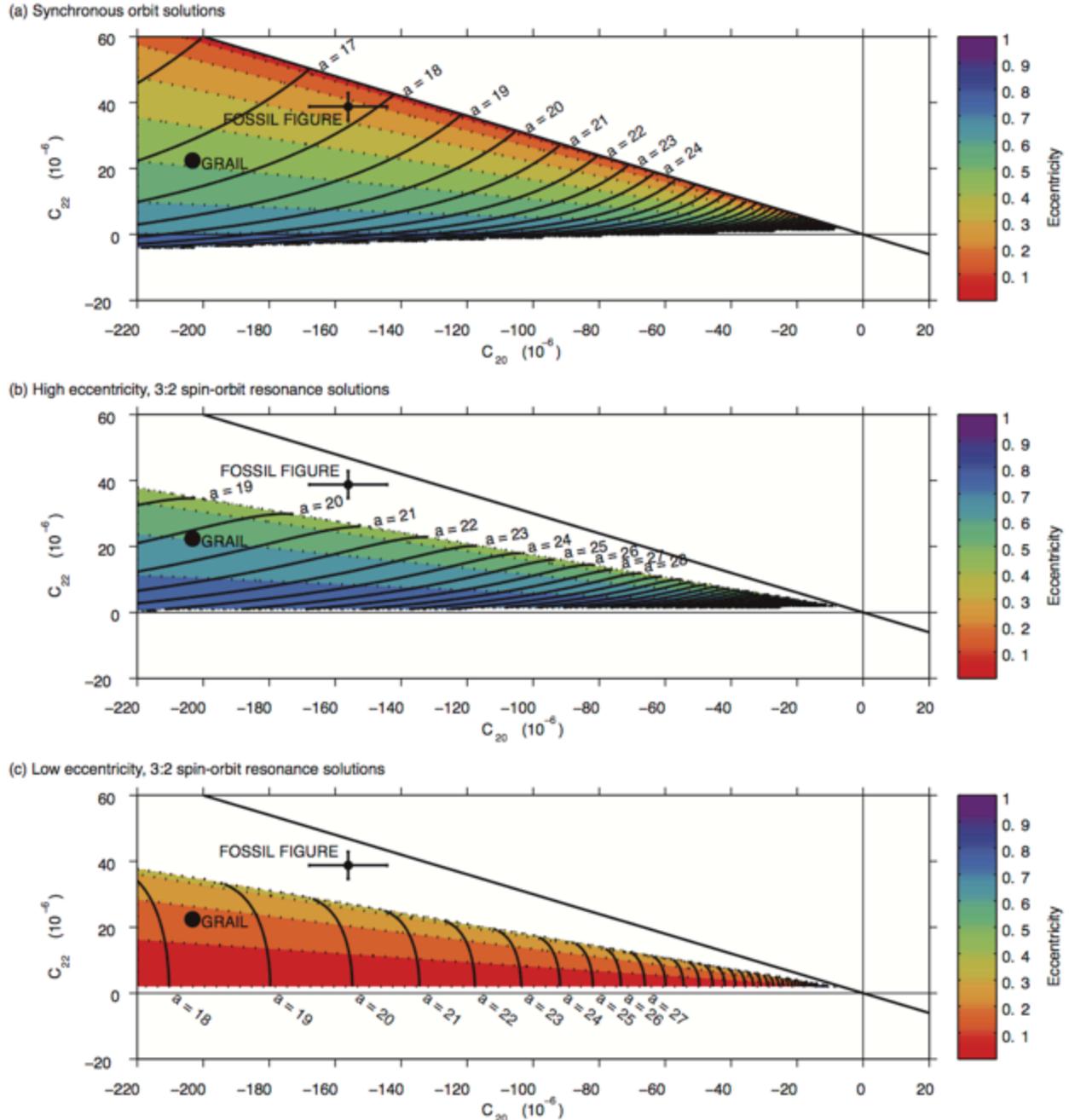

**Figure 2.** Orbital solutions for the mass anomaly and true polar wander corrected fossil figure, assuming a 50 km thick elastic lithosphere, as a function of degree-2 gravity coefficients, $C_{20}$ and $C_{22}$. Colors and dashed contours indicate the eccentricity at the time the elastic lithosphere formed; solid black contours indicate the semimajor axis (in Earth radii) at the time the elastic lithosphere formed. The diagonal black line corresponds to the classical, zero-eccentricity, hydrostatic solution ($C_{22}/C_{20}=-3/10$). White regions correspond to parameter space where no self-consistent solution exists. The black circle indicates the observed GRAIL degree-2 gravity coefficients. The black point with 1σ error bars indicates the mass anomaly and true polar wander corrected fossil figure that results after subtracting our best fit spherical cap model from the observed degree-2 coefficients and diagonalizing. (a) Solutions for a 1:1 (synchronous) spin-orbit resonance. (b and c) High- and low-eccentricity solutions for a 3:2 spin-orbit resonance. There is no self-consistent fossil figure solution for a 3:2 spin-orbit resonance.

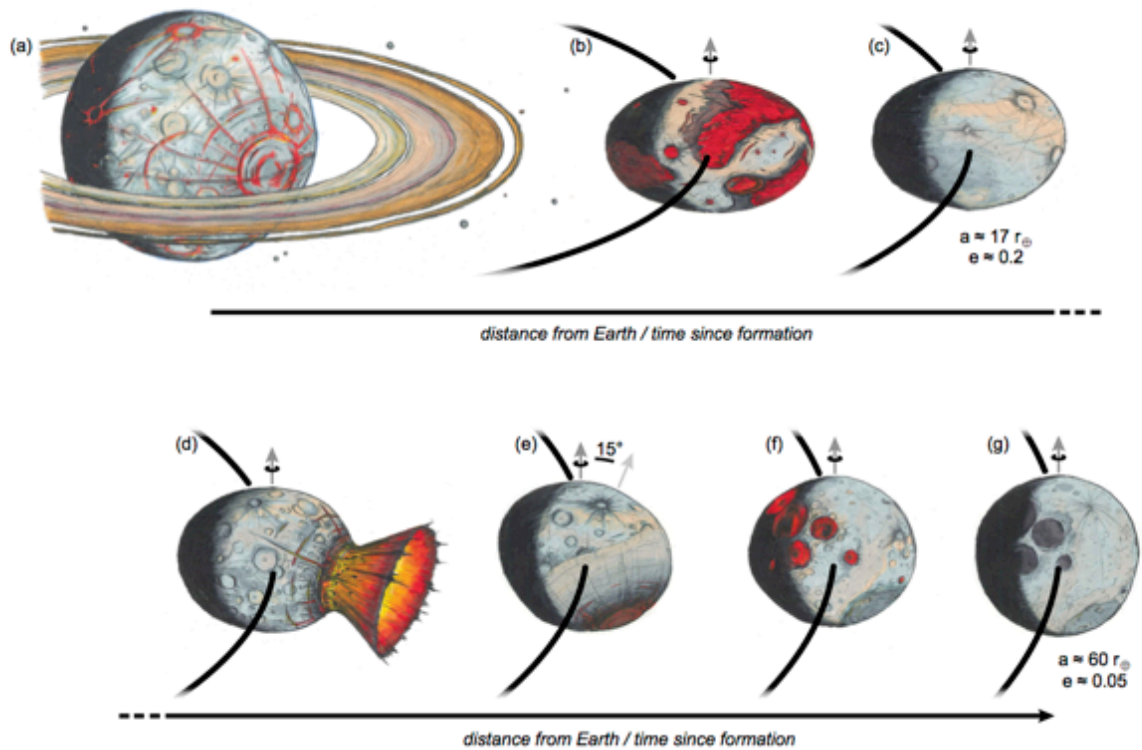

**Figure 3.** Our proposed model for the formation of the Moon's fossil figure. (a) The Moon coalesced from the debris of giant impact. (b) Under the action of tides, the Moon migrated outward and cooled from a magma ocean, perhaps experiencing periods of non-synchronous rotation or high eccentricity. (c) As the moon cools, it forms an elastic lithosphere capable of supporting long-term deformation, resulting in the fossil figure. (d-e) The SPA basin forming impact occurs, resulting in ~15° of reorientation, placing SPA closer to the south pole. (f) Subsequent large impacts and mare volcanism occur, but do not significantly alter the lunar figure. (g) Ultimately, the Moon migrates to its current orbital configuration.

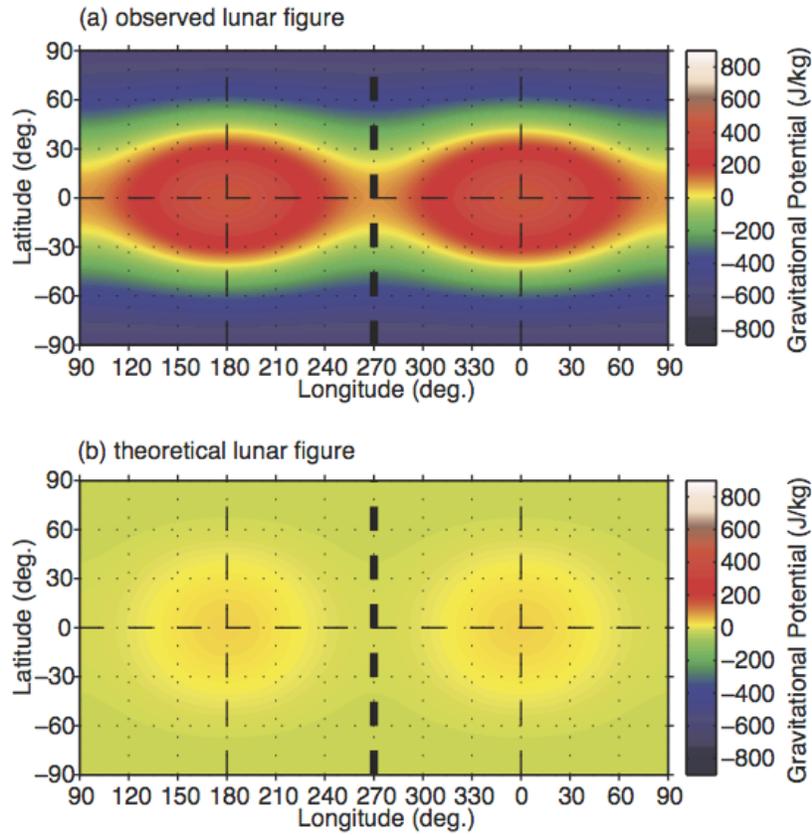

**Figure S1.** The observed and theoretical, hydrostatic degree-2 gravitational potential of the Moon (henceforth, the lunar "figure"). (a) The GRAIL observed lunar figure. (b) The theoretical lunar figure predicted assuming the lunar figure is in hydrostatic equilibrium with its current orbital and rotational state, evaluated using equations (11) and (12).

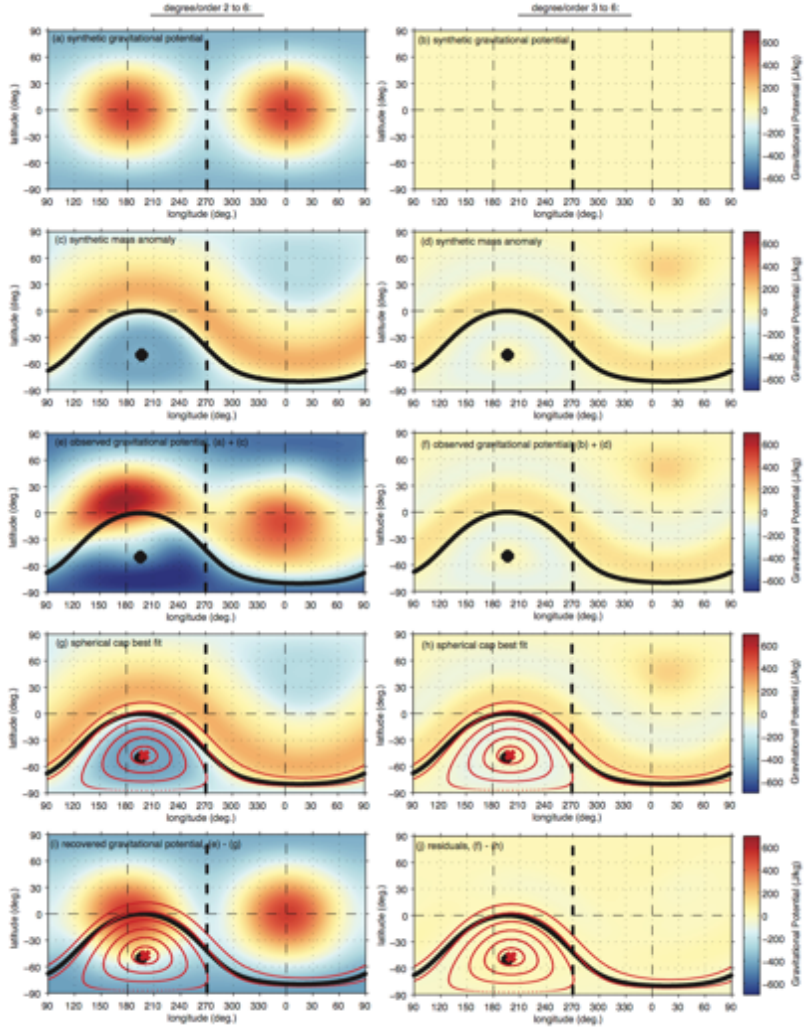

**Figure S2.** Demonstration of the mass anomaly fitting process for a large SPA-like mass anomaly, by way of the forward problem. Panels on the left display the gravitational potential evaluated from spherical harmonic degree 2 to 6, while panels on the left show the gravitational potential evaluated from degree 3 to 6. While the former contains information about the inertia tensor (in degree-2), the latter is important, as it is the dataset that we are actually using in the least squares fitting. To demonstrate this method, we first create a synthetic fossil figure comprised solely of degree-2 gravity terms: $C_{20} = -130 \times 10^{-6}$ and $C_{22} = 40 \times 10^{-6}$ (a and b). We then use equations (1) and (2) to create a synthetic mass anomaly comprised of a single spherical cap with a radius of 50° centered on -50° latitude, 196 longitude with a negative surface density equivalent to a 0.5 km deep cavity devoid of 3 g/cc rock (c and d). The center and radial extent of this synthetic mass anomaly are indicated by the thick black cross and contour. This synthetic mass anomaly is meant to emulate the South Pole-Aitken impact basin. Adding this synthetic mass anomaly (c and d) to the synthetic fossil figure (a and b) yields the total, observable figure (e and f). We then attempt to retrieve the synthetic fossil figure by fitting the observed gravitational potential from degree-3 and up (f) with a set of uniform density spherical caps following the methodology described above. The radii of the set of spherical caps are selected to span the full size of the synthetic mass anomaly, but with no radii exactly equal to the size of the synthetic anomaly. The center of the spherical caps is also purposefully misaligned with the center of the synthetic mass anomaly, to simulate the effects of any potential misalignment of mass anomalies. We then perform the inverse operation and determine the best fitting mass anomaly model (g and h). The center and radial extent of the set of spherical caps are indicated by the red cross and red contours. Finally, we subtract the best-fit mass anomaly model from the observed figure, producing the recovered fossil figure (i) and fit residuals (j). Even with a severe 5° induced misalignment between the center of the mass anomaly and the spherical caps used to fit the anomaly, we are able to retrieve $C_{20}$ to within 8% and $C_{22}$ to within 1%.

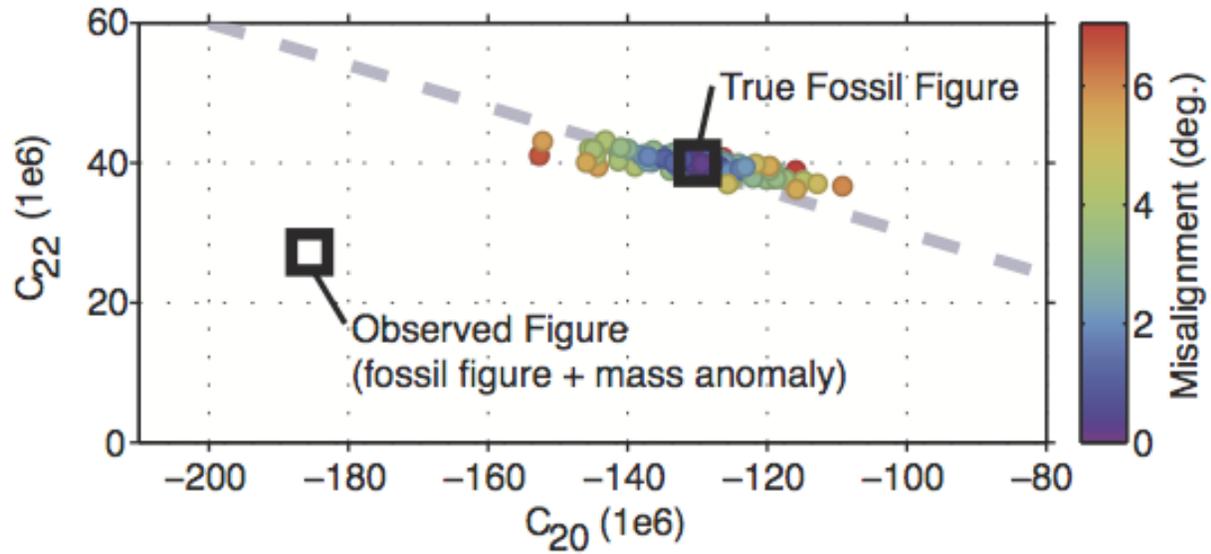

**Figure S3.** Demonstration of the accuracy of the spherical cap inverse method. Like figure S1, we generated a synthetic fossil figure and synthetic SPA-like mass anomaly and then used the spherical cap inverse method to attempt to retrieve the input fossil figure from the observed figure (the sum of the synthetic fossil figure and the synthetic mass anomaly). We repeated the inverse several hundred times, randomly offsetting the position of the spherical caps with respect to the true center of the mass anomaly. Assuming we are able to accurately place the centers of mass anomalies within ~5°, we are able to retrieve $C_{20}$ within 8%, and within 5% for $C_{22}$.

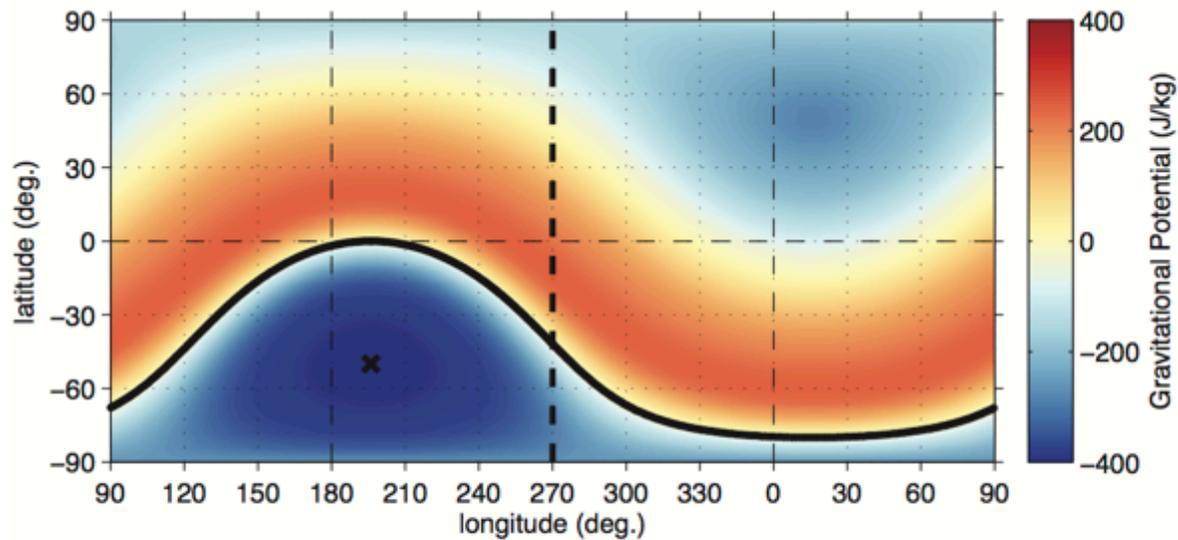

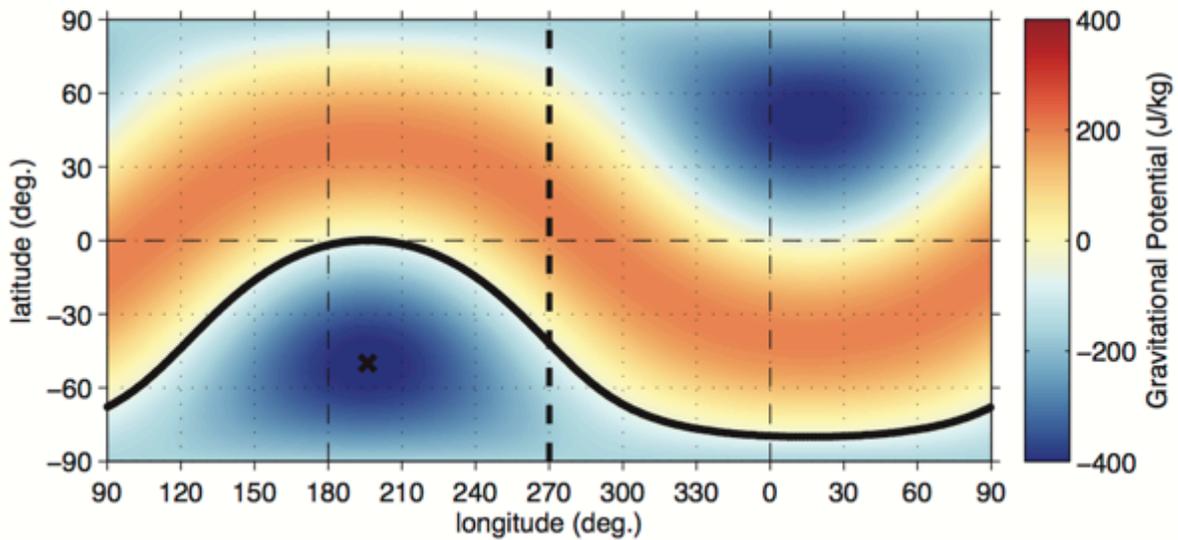

**Figure S4.** The gravitational potential associated with a SPA-like mass anomaly (with no underlying fossil figure; $C_{20} = C_{22} = 0$). (a) The potential evaluated from spherical harmonic degree 2 to 10. (b) The potential evaluated with just spherical harmonic degree 2. While most of the high-order (>3) associated with SPA is concentrated within the actual impact basin, the associated degree-2 field is global. Fitting the degree-2 gravitational potential outside of SPA (as done by Garrick-Bethell et al., 2014) will retain almost all of the degree-2 power associated with SPA.

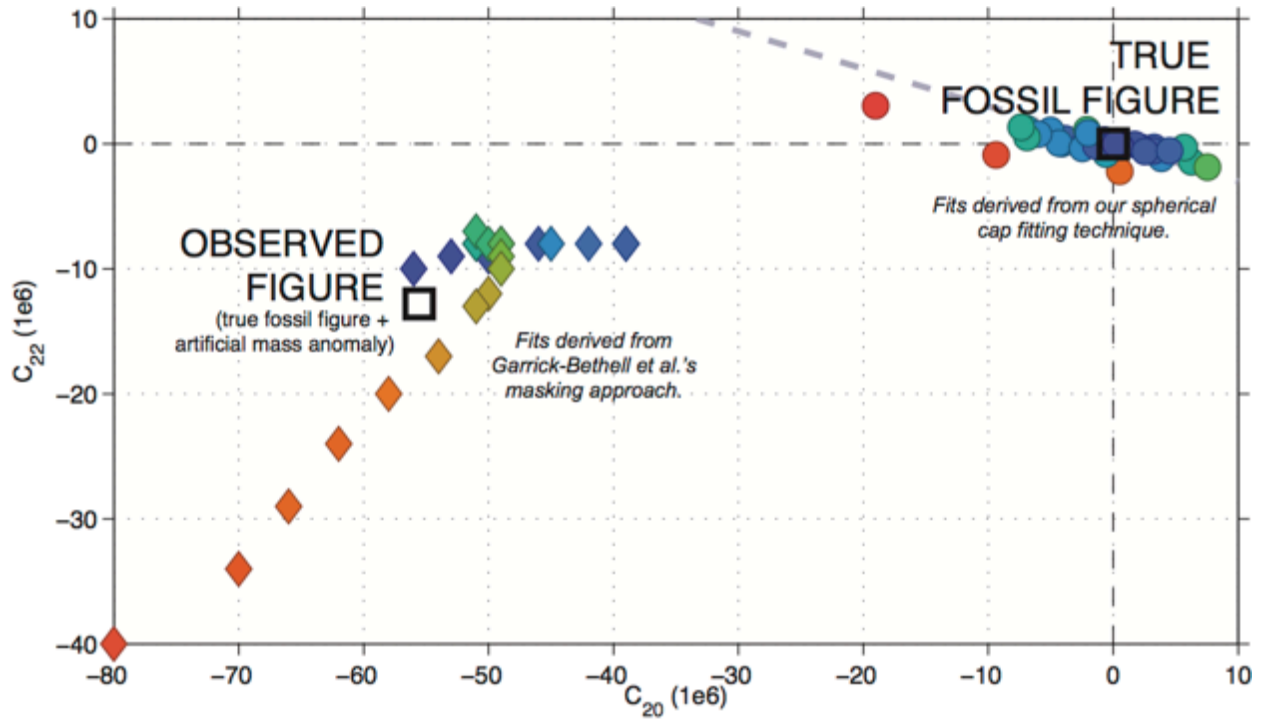

**Figure S5.** A comparison between the spherical cap inverse method used in this work, and the masking approach used by Garrick-Bethell et al. (2014), via the forward problem. For this comparison, we considered the null solution where the Moon possesses no fossil figure ($C_{20} = C_{22} = 0$), but possesses an SPA-like mass anomaly (a single spherical cap with a radius of 50° centered on -50° latitude, 196 longitude with a negative surface density equivalent to a 0.5 km deep cavity devoid of 3 g/cc rock, as in figure S2). Circles indicate fossil figure solutions derived using our spherical cap method (as described in the text), while diamonds indicate fossil figure solutions derived by masking the SPA-like mass anomaly and fitting the remaining gravitational potential. The spherical cap inverse solutions cluster around the true fossil figure solution ($C_{20} = C_{22} = 0$), whereas the masking solutions cluster around the observed figure (the true fossil figure plus the SPA-like mass anomaly). *The masking approach is not capable of retrieving the original fossil figure.* Colors associated with the spherical cap solutions correspond to the magnitude of the imposed misalignment between the spherical caps and the mass anomaly (as in figure S2). Colors associated with the masking solutions correspond to the size of the mask, which ranged from 0.5 to 3 times the size of the radius of the mass anomaly. No misalignment uncertainty was considered for the masking approach.

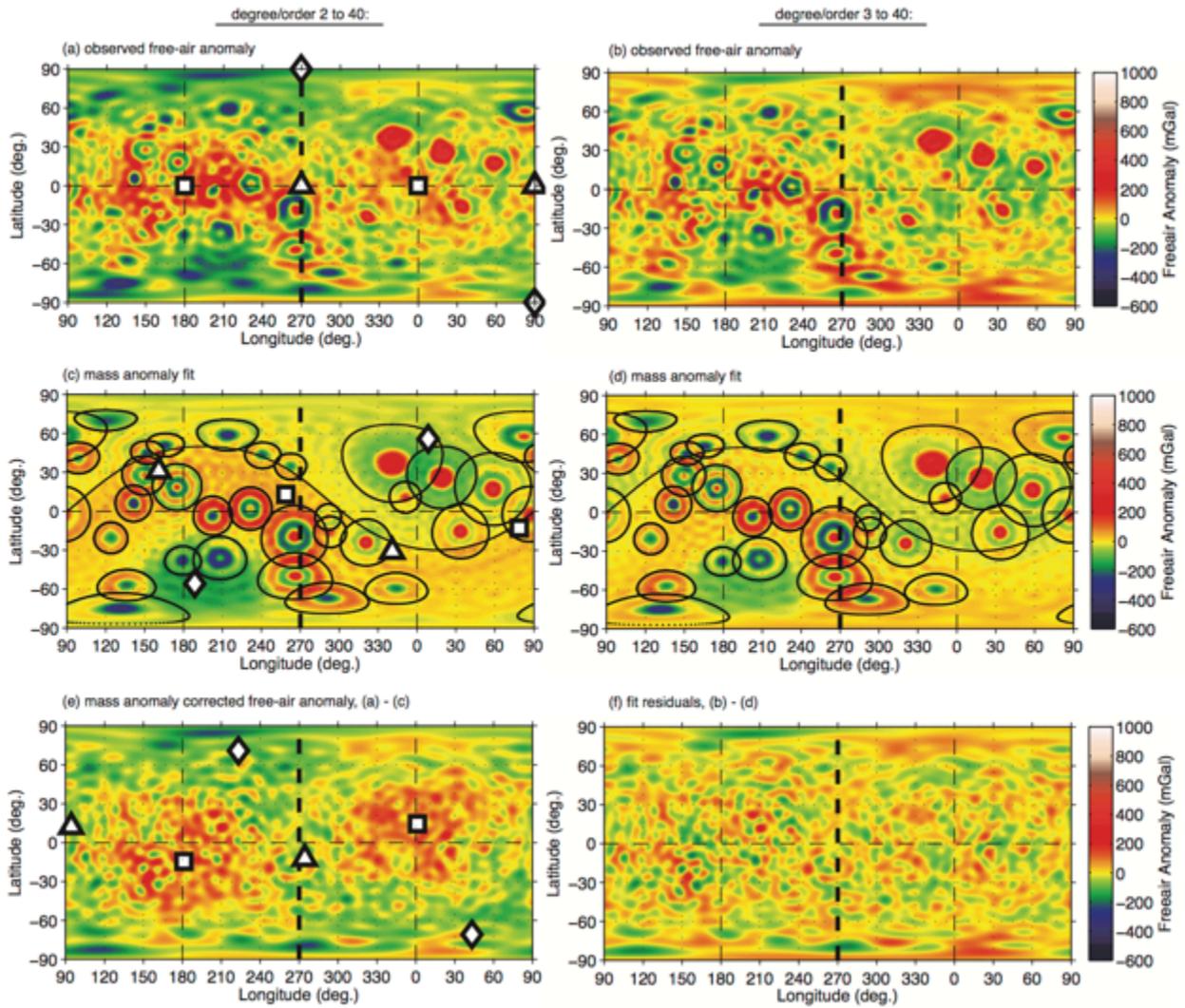

**Figure S6.** The lunar figure, as shown with free-air anomaly. Contours and symbols are the same as figure 1.

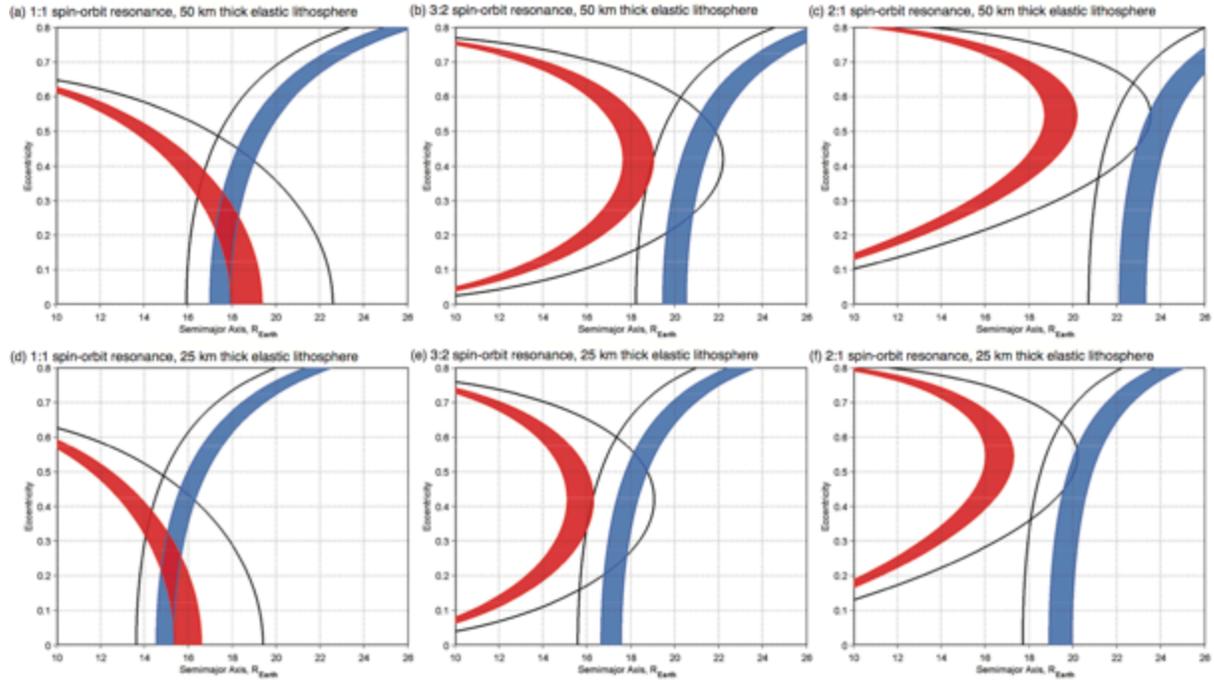

**Figure S7.** Orbital solutions for the mass anomaly and true polar wander corrected fossil figure, plotted in the format of Garrick-Bethell et al. [2006] and Matsuyama [2013]. The blue band indicates the region of orbital parameter space that would yield the appropriate value of $C_{20}$. The red band indicates the region of orbital parameter space that would yield the appropriate value of $C_{22}$. The width of the band corresponds to the $\pm 1\sigma$ uncertainties in $C_{20}$ and $C_{22}$. The region where they overlap (if at all) indicates a region of parameter space that is capable of self-consistently generating the observed mass anomaly and true polar wander corrected fossil figure (corresponding to the black point with error bars in figures 2, S8, and S9). The observed GRAIL values of $C_{20}$ and $C_{22}$ are shown as black contours. Note that the mass anomaly and true polar wander corrected $C_{20}$ and $C_{22}$ constraints only intersect for the case of a 1:1 spin-orbit resonance.

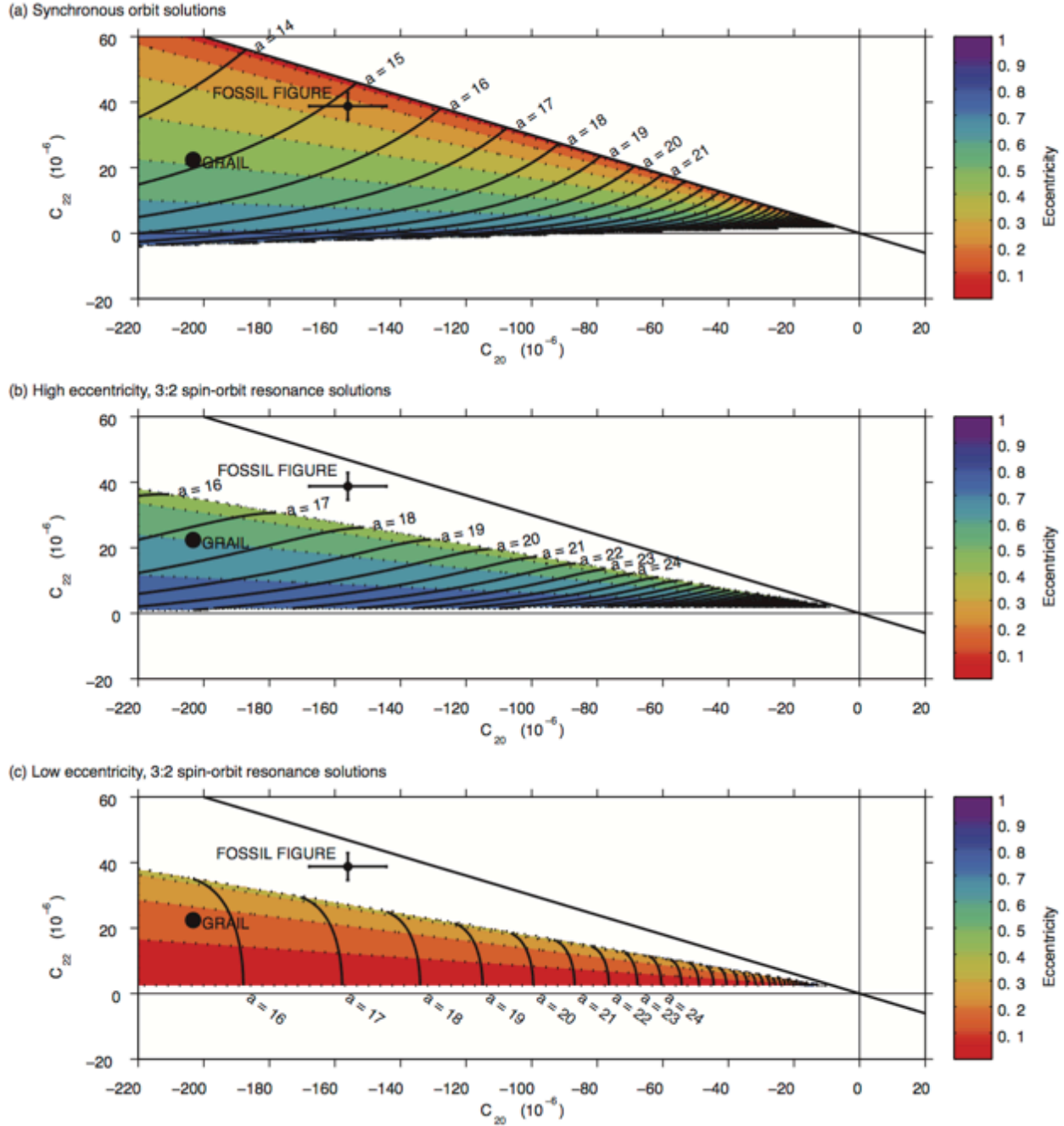

**Figure S8.** 1:1 and 3:2 spin-orbit resonance solutions for the mass anomaly and true polar wander corrected fossil figure, assuming a 25 km thick elastic lithosphere. Contours and symbols are the same as in figure 2.

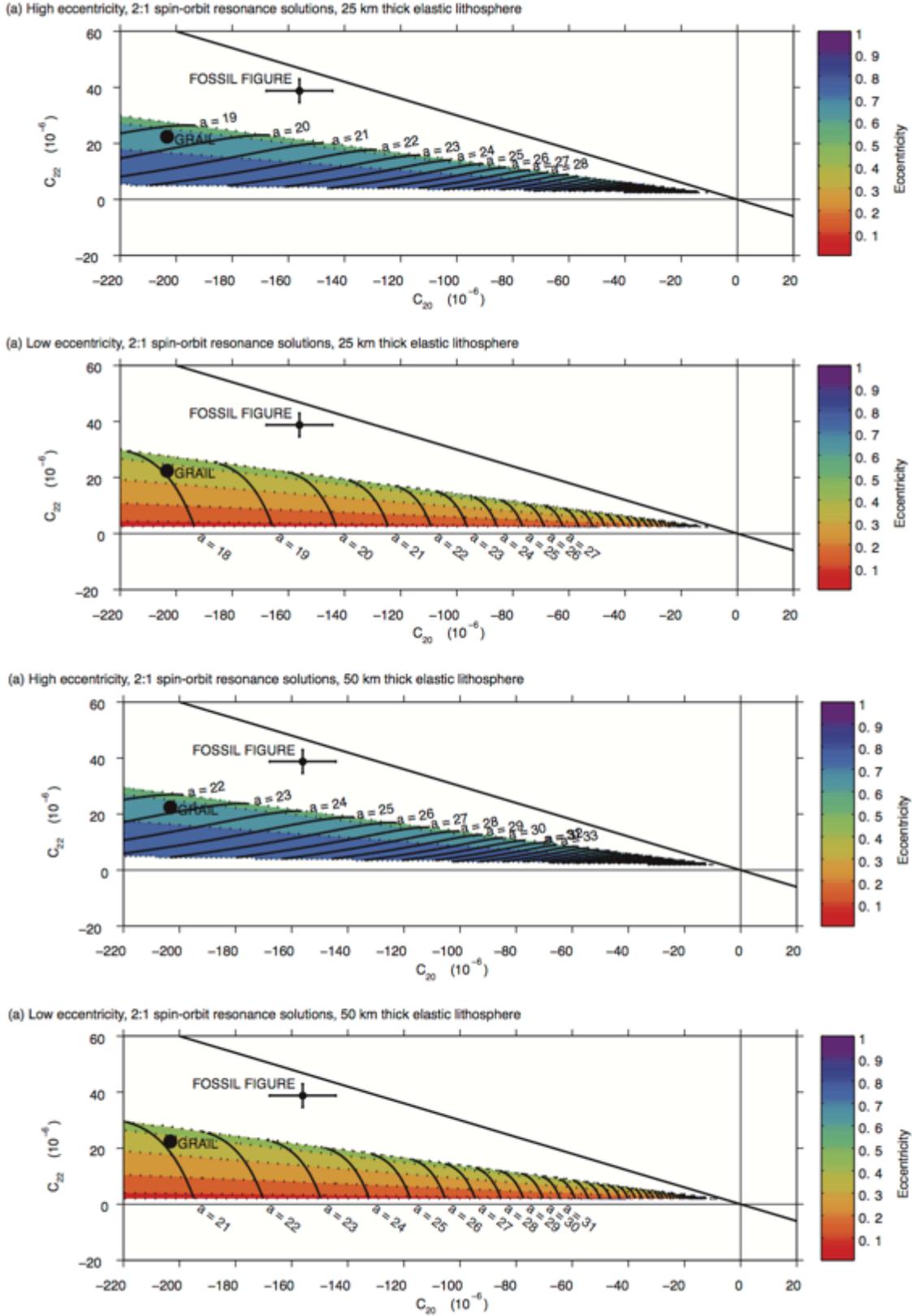

**Figure S9.** 2:1 spin-orbit resonance solutions for the mass anomaly and true polar wander corrected fossil figure. Contours and symbols are the same as in figure 2.

**Table S1**. Second and third order spherical harmonic gravity coefficients (unnormalized).

| Coefficient | GRAIL[A] | All Mass Anomalies[B] | SPA Only[B] | Fossil Figure[B] | Diagonalized Fossil Figure[C] |
|---|---|---|---|---|---|
| | ×10⁻⁶ | × 10⁻⁶ | × 10⁻⁶ | × 10⁻⁶ | × 10⁻⁶ |
| $C_{20}$ | -203.2 | -77.0 | -56.7 | -126.2 | -156.1 ± 11.8 |
| $C_{21}$ | 0.0 | -56.0 | -70.6 | 56.0 | – |
| $C_{22}$ | 22.4 | -16.5 | -13.1 | 38.9 | 38.8 ± 4.1 |
| $S_{21}$ | 0.0 | -15.6 | -20.2 | 15.6 | – |
| $S_{22}$ | 0.0 | 0.3 | -8.2 | -0.3 | – |
| $C_{30}$ | -8.5 | -3.0 | -2.2 | – | – |
| $C_{31}$ | 28.5 | 28.5 | 25.5 | – | – |
| $C_{32}$ | 4.8 | 4.9 | 5.7 | – | – |
| $C_{33}$ | 1.7 | 1.5 | 0.6 | – | – |
| $S_{31}$ | 5.9 | 4.1 | 7.3 | – | – |
| $S_{32}$ | 1.7 | 2.0 | 3.6 | – | – |
| $S_{33}$ | -0.2 | -0.1 | 0.7 | – | – |

[A] GL0420A [Zuber et al., 2013]
[B] Spherical harmonic gravity coefficients from an example fit.
[C] Average and one standard deviation gravity coefficients from the full ensemble of fits.